\documentclass[9pt]{sig-alternate}
\usepackage{subfigure}
\usepackage{graphicx}
\usepackage{cite}
\usepackage{threeparttable}
\usepackage{ntheorem}
\usepackage{color}
\usepackage{multirow}
\usepackage{amsmath}
\usepackage{comment}

\usepackage{algpseudocode}
\usepackage{algorithm}
\theoremstyle{plain}
\theoremheaderfont{\normalfont\bfseries}\theorembodyfont{\slshape}
\theoremsymbol{\ensuremath{\diamondsuit}} \theoremseparator{:}
\newtheorem{mytheorem}{Theorem}

\theoremstyle{plain}
\theoremsymbol{\ensuremath{\clubsuit}}
\theoremseparator{.}
\newtheorem{myproblem}{Problem}

\theoremsymbol{\ensuremath{\heartsuit}}
\newtheorem{mylemma}{Lemma}

\theoremstyle{change}
\theorembodyfont{\upshape}
\theoremsymbol{\ensuremath{\ast}}
\theoremseparator{}

\theoremheaderfont{\sc}\theorembodyfont{\upshape}
\theoremstyle{nonumberplain} \theoremseparator{}
\theoremsymbol{\rule{1ex}{1ex}}

\theoremstyle{plain} \theoremsymbol{\ensuremath{\clubsuit}}
\theoremseparator{.}
\newtheorem{mydefinition}{Definition}

\graphicspath{{./figs/}}

\begin{document}

% ========== following setting is for sig-alternate ============ %
\conferenceinfo{DAC'14,} {June 01--05 2014, San Francisco, CA, USA}
\CopyrightYear{2014} 
\crdata{978-1-4503-2730-5/14/06}

\title{
%\LARGE
Layout Decomposition for Quadruple Patterning \\
Lithography and Beyond
\vspace{-.2in}
%\vspace{.3in}
}

% ==================== ACM style author list ==================== %
\numberofauthors{2}
\author{
\alignauthor Bei Yu\\
  \affaddr{ECE Department}\\
  \affaddr{University of Texas at Austin, Austin, TX}\\
  \email{bei@cerc.utexas.edu}
\alignauthor David Z. Pan\\
  \affaddr{ECE Department}\\
  \affaddr{University of Texas at Austin, Austin, TX}\\
  \email{dpan@ece.utexas.edu}
}

\maketitle
\thispagestyle{empty}%essential

\begin{abstract}

% === shorter version, modified by David 11/22/2013
For next-generation technology nodes, multiple patterning lithography (MPL) has emerged as a key solution,
e.g., triple patterning lithography (TPL) for 14/11nm, and quadruple patterning lithography (QPL) for sub-10nm.
In this paper, we propose a generic and robust layout decomposition framework for QPL, which can be further extended to handle any general K-patterning lithography (K$>$4).
Our framework is based on the semidefinite programming (SDP) formulation with novel coloring encoding.
Meanwhile, we propose fast yet effective coloring assignment and achieve significant speedup.
To our best knowledge, this is the first work on the general multiple patterning lithography layout decomposition.

% === longer version
\iffalse
As the feature size of semiconductor technology nodes continues to scale down,
multiple patterning lithography (MPL) has emerged as a key solution for the next generation lithography,
e.g., triple patterning lithography (TPL) for 14/11nm, and quadruple patterning lithography (QPL) for sub-10nm \cite{LITH_2013_Borodovsky}.
However, little research has been reported to date on the layout design complexity under quadruple patterning.
In this paper we propose a semidefinite programming (SDP) based color assignment methodology,
which is also shown to be robust and generic that it can be extended to general K-patterning layout decomposition.
Moreover, we present a faster color assignment algorithm to achieve further speedup.
The experimental results show that the proposed algorithms are very effective.
To the best of our knowledge, this is the first systematic study on general K-patterning lithography (K$>=$4) layout decomposition.
\fi

\end{abstract}

\iftrue
\vspace{-0.1in}
\category{B.7.2}{Hardware, Integrated Circuit} Design Aids
\vspace{-0.1in}
\terms{Algorithms, Design, Performance}
\vspace{-0.1in}
\vspace{-0.1in}
\fi

\section{Introduction}

% ========================================================
%                 Industry Background
% ========================================================
As the minimum feature size further decreases, multiple patterning lithography (MPL) has become one of the most viable solutions to sub-14nm half-pitch patterning,
along with extreme ultra violet lithography (EUVL), electric beam lithography (EBL), and directed self-assembly (DSA) \cite{LITH_TCAD2013_Pan,LITH_ICCAD2012_Yu}.
Last few years have seen extensive researches on MPL technology such as double patterning \cite{DPL_ICCAD08_Kahng},
and triple patterning \cite{TPL_ICCAD2011_Yu}.
Continuing growth of technology node is expected to shrink further down to 11nm or beyond.
Such advance is, nonetheless, making conventional patterning processes barely sufficient for the next generation.
%Triple patterning lithography and self-aligned double patterning, which are two typical versions of multiple patterning,
%are solution candidates for the 14nm logic node \cite{TPL_SPIE2012_Lucas}.
%In the future, as the technology node shrinks further down to 11nm and beyond, both of them will not be sufficient.
%Therefore, quadruple patterning lithography (QPL) would debut, as a new version of multiple patterning.
%In order to remedy this deficiency, quadruple patterning lithography (QPL) has been recently introduced from industry as a new type of MPL technology \cite{ITRS}.

\begin{figure}[tb]
  \vspace{-.1in}
  \centering
  \includegraphics[width=0.40\textwidth]{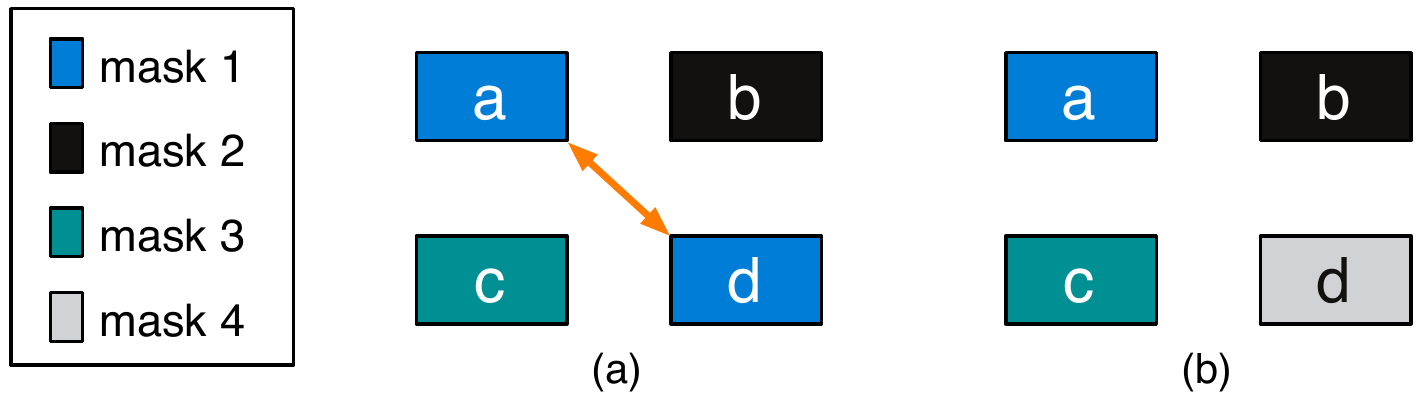}
  \caption{~(a) A common native conflict from triple patterning lithography;~(b) The conflict can be resolved through quadruple patterning lithography.}
  \label{fig:TPL2QPL}
  \vspace{-.2in}
\end{figure}

% =========================================================
%                    QPL is practical
% =========================================================
Quadruple patterning lithography (QPL) is a natural extension along the paradigm of double/triple patterning.
In the QPL manufacturing, there are four exposure/etching processes, through which the initial layout can be produced.
Compared with triple patterning lithography, QPL introduces one more mask.
Although increasing the number of processing steps by 33\% over triple patterning, there are several reasons/advantages for QPL.
Firstly, due to the delay or uncertainty of other lithography techniques, such as EUVL,
semiconductor industry needs CAD tools to be prepared and understand the complexity/implication of QPL.
Even from theoretical perspective, studying the general multiple patterning is valuable.
Secondly, it is observed that for triple patterning lithography, even with stitch insertion, there are several common native conflict patterns.
As shown in Fig. \ref{fig:TPL2QPL} (a), contact layout within the standard cell may generate some 4-clique patterns, which are indecomposable.
This conflict can be easily resolved if four masks are available (see Fig. \ref{fig:TPL2QPL} (b)).
Thirdly, with one more mask, some stitches may be avoided during manufacturing.
By this way it is potential to resolve the overlapping and yield issues derived from the stitches.

% ====================================================================
%                     Layout Decomposition
% ====================================================================
The process of QPL brings up several critical yet open design challenges, such as layout decomposition,
where the original layout is divided into four masks (colors).
%During the layout decomposition, when the distance between two patterns is less than minimum colorable distance $min_s$,
%they need to be assigned to different masks to avoid a \textit{conflict}.
%Sometimes conflict can be solved by splitting a feature into two touching parts, called \textit{stitch}.
%However, since stitch may cause yield loss due to the overlapping issue,
%the two objectives of layout decomposition are to minimize the conflict and the stitch number.
Double/triple patterning layout decomposition with conflict and stitch minimization has been well studied for full-chip layout
\cite{DPL_ICCAD08_Kahng,DPL_ISPD09_Yuan,DPL_ISPD2010_Xu,DPL_ICCAD2011_Tang,
%DPL_ASPDAC2010_Yang,
%TPL_SPIE08_Cork,TPL_ISQED2013_Chen}.
TPL_ICCAD2011_Yu,TPL_DAC2012_Fang,TPL_DAC2013_Kuang,TPL_ICCAD2013_Yu,TPL_ICCAD2013_Zhang,TPLEC_SPIE2013_Yu}
and cell based design
\cite{TPL_ICCAD2012_Tian,DFM_ICCAD2013_Yu,TPL_ICCAD2013_Tian}.
The problem can be optimally solved through expensive integer linear programming (ILP) \cite{DPL_ICCAD08_Kahng,DPL_ISPD09_Yuan,TPL_ICCAD2011_Yu}.
To overcome the long runtime problem of ILP solver, for double patterning, partitioning/matching based methods have been proposed
\cite{DPL_ISPD2010_Xu,DPL_ICCAD2011_Tang};
%DPL_ASPDAC2010_Yang,
while for triple patterning, some speedup techniques, e.g.,
semidefinite programming (SDP) \cite{TPL_ICCAD2011_Yu,TPL_ICCAD2013_Yu},
and heuristic coloring assignment \cite{TPL_DAC2012_Fang,TPL_DAC2013_Kuang}
%and partitioning \cite{TPL_ISQED2013_Chen},
have been proposed.
However, how to effectively solve the quadruple patterning, or even general multiple patterning problems, is still an open question.

% ============    Our contributions    ============= %
In this paper, we deal with the quadruple patterning layout decomposition (QPLD) problem.
Our contributions are highlighted as follows.
(1) To our best knowledge, this is the first layout decomposition research for QPLD problem.
We believe this work will invoke more future research into this field thereby promoting the scaling of technology node.
(2) Our framework consists of holistic algorithmic processes,
such as semidefinite programming based algorithm, linear color assignment, and novel GH-tree based graph division.
(3) We demonstrate the viability of our algorithm to suit with general K-patterning (K$\ge$4) layout decomposition,
which could be advanced guidelines for future technology. %\cite{LITH_2013_Borodovsky}.

%In this paper, we propose a systematic framework for quadruple patterning layout decomposition.
%To our best knowledge, it is the first layout decomposition study for quadruple patterning lithography.
%First we present the mathematical formulation of the problem.
%Then a set of algorithms, i.e., integer linear programming, heuristic partitioning, and semidefinite programming based algorithm, are proposed to solve the problem.
%To further reduce the problem size, several novel and effective graph simplification techniques are proposed.
%In addition, our framework is robust that it can be extended to handle general multiple patterning layout decomposition.

The rest of the paper is organized as follows.
In Section \ref{sec:prelim}, we give the problem formulations and the overall decomposition flow.
In Section \ref{sec:algo} and Section \ref{sec:graph} we propose the color assignment algorithms and graph division techniques, respectively.
Section \ref{sec:extend} extends our methodologies to general K-patterning problem.
Section \ref{sec:result} presents the experiment results, followed by conclusion in Section \ref{sec:conclu}.

\section{Preliminaries}
\label{sec:prelim}

%Preliminaries of QPLD are introduced in this section,
%including formal problem formulation and our proposed decomposition framework.

\subsection{Problem Formulation}

Given input layout which is specified by features in polygonal shapes, a \textit{decomposition graphs}
\cite{DPL_ISPD09_Yuan,TPL_ICCAD2011_Yu} is constructed by Definition \ref{def:dg}.

\begin{mydefinition}[Decomposition Graph]
\label{def:dg}
A decomposition \\ graph is an undirected graph $\{V, CE, SE\}$ with a single set of vertices $V$,
and two edge sets $CE$ and $SE$ containing the \textit{conflict edges} (CE) and \textit{stitch edges} (SE), respectively.
Each vertex  $v \in V$ represents a polygonal shape, an edge $e \in CE$ exists iff the two polygonal shapes are within minimum coloring distance $min_s$,
and an edge $e \in SE$ iff there is a stitch candidate between the two vertices which are associated with the same polygonal shape.
\end{mydefinition}

% ===============================================================
%       Following parts are skipped for Journal version
% ===============================================================
\iffalse
\begin{figure}[tb]
  \centering
  \subfigure[] {\includegraphics[width=0.14\textwidth]{MPL_LG2DG1}}
  \hspace{.2in}
  \subfigure[] {\includegraphics[width=0.15\textwidth]{MPL_LG2DG2}}
  \caption{~(a) Input layout and all stitch candidates;~(b) Decomposition graph.}
  \label{fig:LG2DG}
  \vspace{-.1in}
\end{figure}

We use Fig. \ref{fig:LG2DG} for illustration.
Firstly, vertex projection is performed on input layout to search all stitch candidates.
Then decomposition graph (see Fig. \ref{fig:LG2DG} (b)) is constructed to store all the layout information and the stitch candidates.
Note that in vertex projection we choose all the possible stitch candidates as described in \cite{TPL_DAC2013_Kuang}.
%Based on decomposition graph, we carry out color assignment to assign all the vertices into four masks (colors).
\fi

Now we give the problem formulation of quadruple patterning layout decomposition (QPLD).

\begin{myproblem}[QPLD]
Given an input layout which is specified by features in polygonal shapes and minimum coloring distance $min_s$,
the decomposition graph is constructed.
Quadruple patterning layout decomposition (QPLD) assigns all the vertices into one of four colors (masks) to minimize conflict number and stitch number.
\end{myproblem}

The QPLD problem can be extended to general K-patterning layout decomposition problem as follows.

\begin{myproblem}[K-Patterning Layout Decomposition]\ \\
Given an input layout, the decomposition graph is constructed.
Each vertex in graph would be assigned into one of K colors (masks) to minimize conflict number and stitch number.
\end{myproblem}

% =================================================================
%                         NP-hardness
% =================================================================
\iffalse
At the first glance, QPLD problem can be solved through \textit{four color map theorem}\cite{1977Appel} that every planar graph is 4-colorable.
However, we observe that in emerging technology node, the designs are so complex that for most cases the layout graphs are not planar.
In addition, we have the following lemma that inserting stitch cannot planarize the graph.

\begin{mylemma}
\label{lem:planar}
If a layout graph (LG) is not planar, after stitch candidate insertion, the corresponding decomposition graph is still not planar.
\end{mylemma}

The proof can be based on \textit{Kuratowski's theorem} \cite{1930Kuratowski} that
a finite graph is planar iff it does not contain a subgraph that is a subdivision of $K_5$ or of $K_{3,3}$.
Here $K_5$ is the complete graph on five vertices, while $K_{3,3}$ is the complete bipartite graph on six vertices (three on each side).
Due to space limit, the detailed proof is omitted.

It has been shown that triple patterning layout decomposition, which is a $3$ patterning problem, is NP-hard \cite{TPL_ICCAD2011_Yu}.
Besides, since triple patterning layout decomposition can be polynomial-time reducible to QPLD,
QPLD problem is NP-hard as well.
In addition, we can get the conclusion that general K-patterning layout decomposition problem, with $K \ge 3$, is NP-hard.
\fi

\subsection{Overview of Layout Decomposition Flow}

\begin{figure}[htb]
  \vspace{-.1in}
  \centering
  \includegraphics[width=0.40\textwidth]{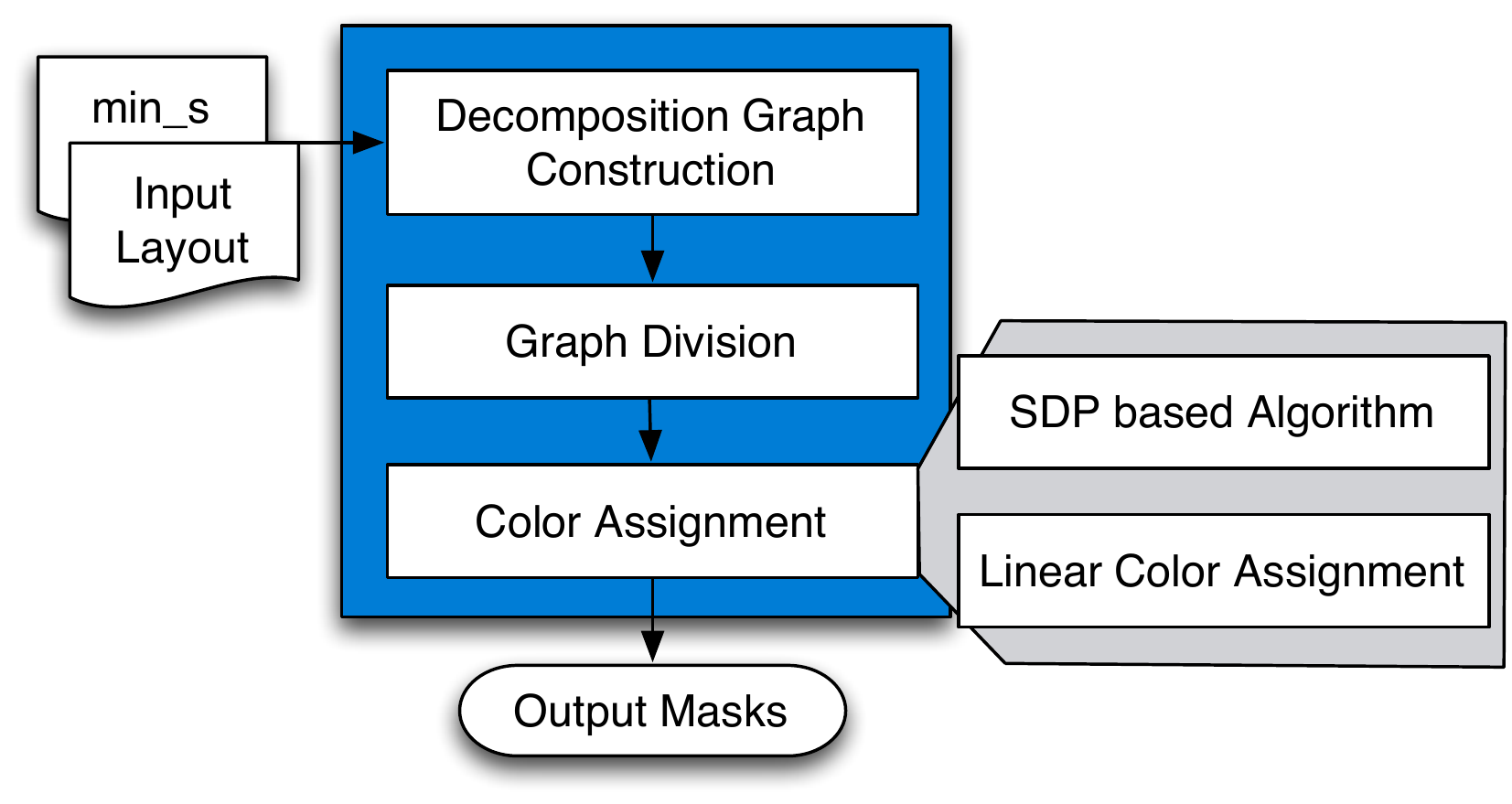}
  \caption{~Proposed layout decomposition flow.}
  \label{fig:overview}
  \vspace{-.1in}
\end{figure}

The overall flow of our layout decomposition is summarized in Fig. \ref{fig:overview}.
We first construct decomposition graph to transform the original geometric patterns into a graph model.
By this way, the QPLD problem can be formulated as 4 coloring on the decomposition graph.
To reduce the problem size, graph division techniques (see Section \ref{sec:graph}) are applied to partition the graph into a set of components.
Then the color assignment problem can be solved independently for each component, through a set of algorithms discussed in Section \ref{sec:algo}.
%i.e., integer linear programming (ILP), semidefinite programming (SDP) based algorithm, and fast color assignment.

\vspace{-.1in}
\section{Color Assignment in QPLD}
\label{sec:algo}

%After graph division, the whole decomposition graph is transferred into a set of components with each corresponding to an independent subproblem.
Given decomposition graph $G = \{V, CE, SE\}$, color assignment would be carried out to assign each vertex into one of four colors (masks),
to minimize both the conflict number and the stitch number.
In this section, we propose two color assignment algorithms, i.e.,
semidefinite programming (SDP) based algorithm, and linear color assignment.
%we first give the mathematical formulation for the color assignment in QPLD,

% ============================================================
%           Subsection: SDP based Color Assignment
% ============================================================

\subsection{SDP Based Color Assignment}
\label{sec:sdp}

\begin{figure}[htb]
  \vspace{-.1in}
  \centering
  \includegraphics[width=0.30\textwidth]{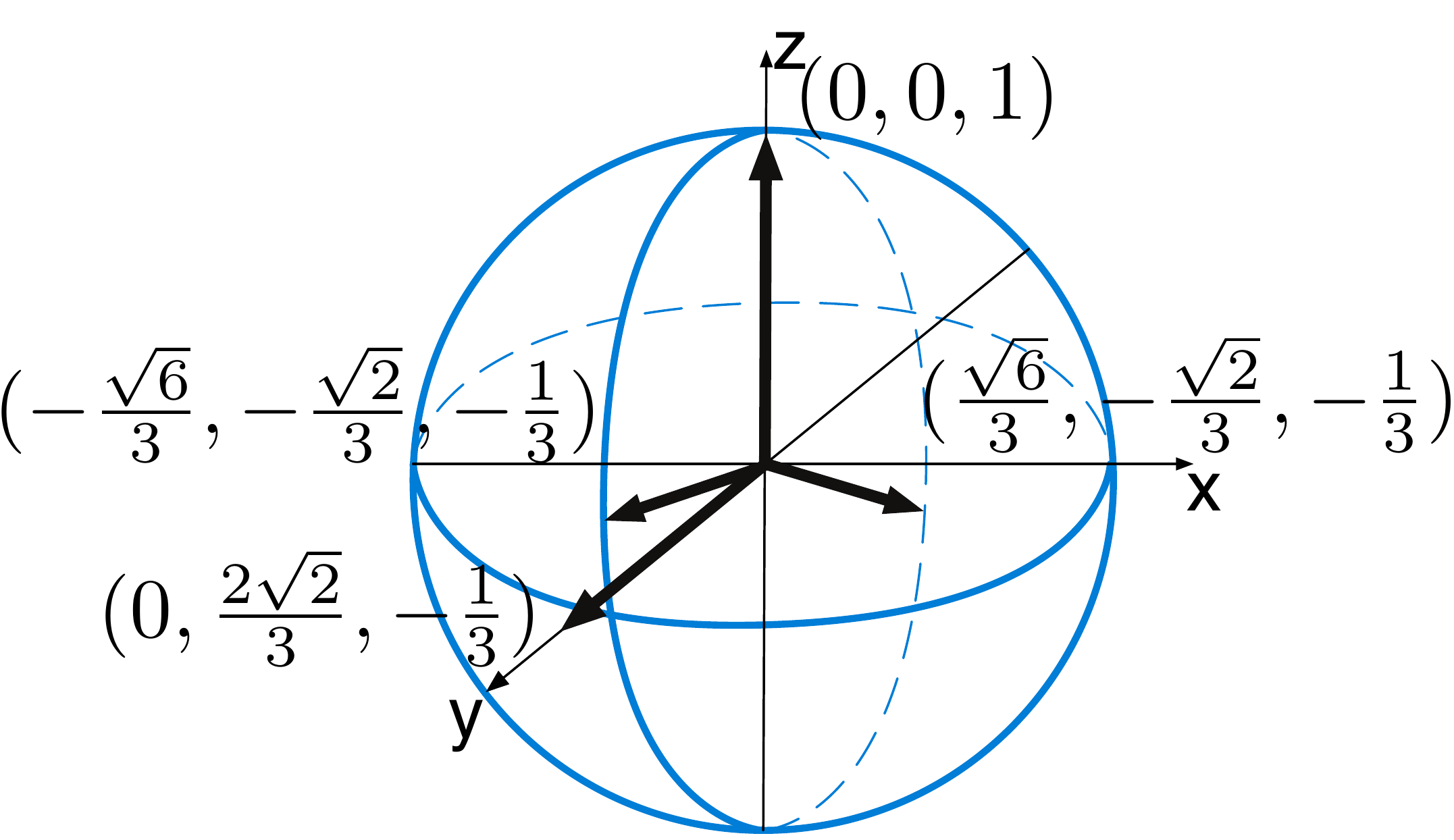}
  \caption{~Four vectors correspond to four colors.}
  \label{fig:vector}
  \vspace{-.1in}
\end{figure}

Semidefinite programming (SDP) has been successfully applied to triple patterning layout decomposition \cite{TPL_ICCAD2011_Yu,TPL_ICCAD2013_Yu}.
Here we will show that SDP formulation can be extended to solve QPLD problem.
%and we will show that it can be extended to general $k$ patterning layout decomposition.
%Therefore, the SDP based TPL layout decomposer \cite{TPL_ICCAD2011_Yu}\cite{TPL_ICCAD2013_Yu} can be re-used.
To represent four different colors (masks), as illustrated in Fig. \ref{fig:vector}, four unit vectors are introduced \cite{1998Karger}:
$(0, 0, 1)$, $(0, \frac{2\sqrt{2}}{3}, -\frac{1}{3})$, $(\frac{\sqrt{6}}{3}, -\frac{\sqrt{2}}{3}, -\frac{1}{3})$ and $(-\frac{\sqrt{6}}{3}, -\frac{\sqrt{2}}{3}, -\frac{1}{3})$.
%Note that the vector is unit vector, and the dimension is 3 now for quadruple patterning.
We construct the vectors in such a way that inner product for any two vectors $\vec{v_i}, \vec{v_j}$ satisfying: 
$\vec{v_i} \cdot \vec{v_j} = 1$ if $\vec{v_i}=\vec{v_j}$; $\vec{v_i} \cdot \vec{v_j} = -\frac{1}{3} $ if $\vec{v_i} \ne \vec{v_j}$.

Based on the vector definition, the QPLD problem can be formulated as the following vector programming:
\begin{align}
  \label{eq:vp}
  \textrm{min}  & \sum_{e_{ij} \in CE} \frac{3}{4} (\vec{v_i} \cdot \vec{v_j}+\frac{1}{3}) + \frac{3\alpha}{4} \cdot \sum_{e_{ij} \in SE} ( 1 - \vec{v_i} \cdot \vec{v_j} )\\
  \textrm{s.t}. \ \
    & \vec{v_i} \in \{ (0, 0, 1), (0, \frac{2\sqrt{2}}{3}, -\frac{1}{3}), (\frac{\sqrt{6}}{3}, -\frac{\sqrt{2}}{3}, -\frac{1}{3}), \notag\\
    &                  \qquad (-\frac{\sqrt{6}}{3}, -\frac{\sqrt{2}}{3}, -\frac{1}{3})\} \notag
\end{align}
where the objective function is to minimize the conflict number and the stitch number.
$\alpha$ is a user-defined parameter, which is set as $0.1$ in this work.
After relaxing the discrete constraints in (\ref{eq:vp}) and removing the constant in objective function,
we redraw the following semidefinite programming (SDP) formulation.

\begin{align}
  \vspace{-.2in}
  \label{eq:sdp}
  \textrm{min}  & \sum_{e_{ij} \in CE} \vec{v_i} \cdot \vec{v_j} - \alpha \sum_{e_{ij} \in SE} \vec{v_i} \cdot \vec{v_j} \\
  \textrm{s.t}.\ \	
    & \vec{v_i} \cdot \vec{v_i} = 1 ,               \ \ \ \forall i \in V			        \notag\\
    & \vec{v_i} \cdot \vec{v_j} \ge -\frac{1}{3},   \ \ \ \forall e_{ij} \in CE      \notag
  \vspace{.1in}
\end{align}

After solving the SDP, we get a set of continuous solutions in matrix $X$,
where each value $x_{ij}$ in matrix $X$ corresponds to $v_i \cdot v_j$.
%Instead of solving $v_i$ and $v_j$ directly, we pay attention to $v_i \cdot v_j$ value itself.
If $x_{ij}$ is close to $1$, vertices $v_i, v_j$ are tend to be in the same mask (color).
A greedy mapping algorithm \cite{TPL_ICCAD2011_Yu} can be directly applied here to get color assignment solution.
However, the performance of greedy method may not be good.

\begin{algorithm}[h]
\caption{SDP + Backtrack}
\label{alg:backtrack}
%{{{
\begin{algorithmic}[1]
  \Require SDP solution $x_{ij}$, threshold value $t_{th}$;
  \ForAll {$x_{ij} \ge t_{th}$}
      \State Combine vertices $v_i, v_j$ into one larger vertex;
  \EndFor
  \State Construct merged graph $G'=\{V', CE', SE'\}$;
  %\State BRANCH-AND-BOUND($0, G'$);
  \State BACKTRACK($0, G'$);
  \State \Return {color assignment result in $G'$;}

  \Statex
  \Function{BACKTRACK}{$t, G'$}
    \If {t $\ge \textrm{size}[G']$}
      \If { Find a better color assignment }
        \State Store current color assignment;
      \EndIf
    %\ElsIf {LOWER-BOUND( ) $> minCost$}
    %  \State return;
    \Else 
      \ForAll {legal color $c$};
        \State $G'[t] \leftarrow c$;
        \State BACKTRACK($t+1, G'$);
        \State $G'[t] \leftarrow -1$;
      \EndFor
    \EndIf
  \EndFunction

\end{algorithmic}
%}}}
\end{algorithm}

To overcome the limitation of the greedy method,
in our framework a backtrack based algorithm (see Algorithm \ref{alg:backtrack}) is proposed to consider both SDP results and graph information.
%Similar to the work in \cite{TPL_ICCAD2011_Yu} the mapping procedure is required after the SDP formulation to finally determine layout decomposition results.
The backtrack based method accepts two arguments of the SDP solution $\{x_{ij}\}$ and a threshold value $t_{th}$.
In our work $t_{th}$ is set as $0.9$.
%We first formulate the SDP (\ref{eq:sdp}) and solve it to get a solution matrix where all $v_i \cdot v_j$ pairs are computed. 
%Given the decomposition graph $G = \{V, CE, SE\}$, at the beginning the SDP formulation in (\ref{eq:sdp}) is formulated.
%After solving the SDP formulation, we get a matrix, where all $v_i \cdot v_j$ pairs are calculated.
As discussed above, if $x_{ij}$ is close to be 1, two vertices $v_i$ and $v_j$ tend to be in the same color (mask).
Therefore, we scan all pairs, and combine some vertices into one larger vertex (lines $1-3$).
After the combination, the vertex number can be reduced, thus the graph has be simplified (line $4$).
The simplified graph is called \textit{merged graph} \cite{TPL_ICCAD2013_Yu}.
On the merged graph, \textit{BACKTRACK} algorithm is presented to search an optimal color assignment (lines $7-19$).
%It shall be noted that as the problem size has been reduced by vertex merging (line $2$),
%backtrack can be accomplished without serious runtime overhead.

% ============================================================
%           Subsection: Linear Color Assignment
% ============================================================

\subsection{Linear Color Assignment}
\label{sec:linear}

Backtrack based method may still involve runtime overhead, especially for complex case where SDP solution cannot provide enough merging candidates.
Therefore, an efficient color assignment is required.
At first glance, the color assignment for quadruple patterning can be solved through \textit{four color map theorem} \cite{1977Appel} that every planar graph is 4-colorable.
However, in emerging technology node, the designs are so complex that we observe many $K_5$ or $K_{3,3}$ structures,
where $K_5$ is the complete graph on five vertices, while $K_{3,3}$ is the complete bipartite graph on six vertices.
%(three on each side).
Due to \textit{Kuratowski's theorem} \cite{1930Kuratowski}, the decomposition graph is not planar,
thus classical four coloring techniques \cite{1996Robertson} is hard to be applied.

Here we propose an efficient color assignment algorithm.
Note that our method is targeting general graph, not just planar graph.
In addition, different from classical four coloring method that needs quadratic runtime \cite{1996Robertson},
our color assignment is a linear runtime algorithm.

\begin{algorithm}[tb]
\caption{Linear Color Assignment}
\label{alg:linear}
%{{{
\begin{algorithmic}[1]
  \Require Decomposition graph $G=\{V, CE, SE\}$, Stack $S$;
  \While{ $\exists v_i \in V$ s.t. $d_{conf}(v_i)<4$ \& $d_{stit}(v_i) < 2$}
      \State $S$.push($v_i$);
      \State $G$.delete($v_i$);
  \EndWhile
  \State Construct vector $vec$;
  %\State Construct vector $vec$ = \{$vec$[1], $vec$[2], $vec$[3]\};
  \State C1 = SEQUENCE-COLORING($vec$);
  \State C2 = DEGREE-COLORING($vec$);
  \State C3 = 3ROUND-COLORING($vec$);
  \State C = best coloring solution among \{C1, C2, C3\};
  \State POST-REFINEMENT($vec$);
  %\ForAll{edge $e_{ij} \in CE$}                \Comment{post-refinement}
  %  \State Try swap colors $c(v_{i})$ and $c(v_{j})$;
  %\EndFor
  \While{ $! S$.empty()}
  	\State $v_i = S$.pop();
   \State $G$.add($v_i$);
   \State $c(v_i) \leftarrow $ a legal color;
  \EndWhile
\end{algorithmic}
%}}}
\end{algorithm}

The details of linear color assignment is summarized in Algorithm \ref{alg:linear}, which involves three stages.
The first stage is iteratively vertex removal.
For each vertex $v_i$, we denote its conflict degree $d_{conf}(v_i)$ as number of conflict edges incident to $v_i$,
while its stitch degree $d_{stit}(v_i)$ as number of stitch edges.
The main idea is that the vertices with conflict degree less than 4 and stitch degree less than 2 are identified as non-critical,
thus can be temporarily removed and pushed into stack $S$ (lines 1-4).
After coloring remaining vertices, each vertex in stack $S$ would be pop up one by one and assigned one legal color (lines 11-15).
This strategy is safe in terms of conflict number.
In other words, when a vertex is pop up from $S$, there is always one color available without introducing new conflict.

\begin{figure}[tb]
  \centering
  \subfigure[]{\includegraphics[width=0.14\textwidth]{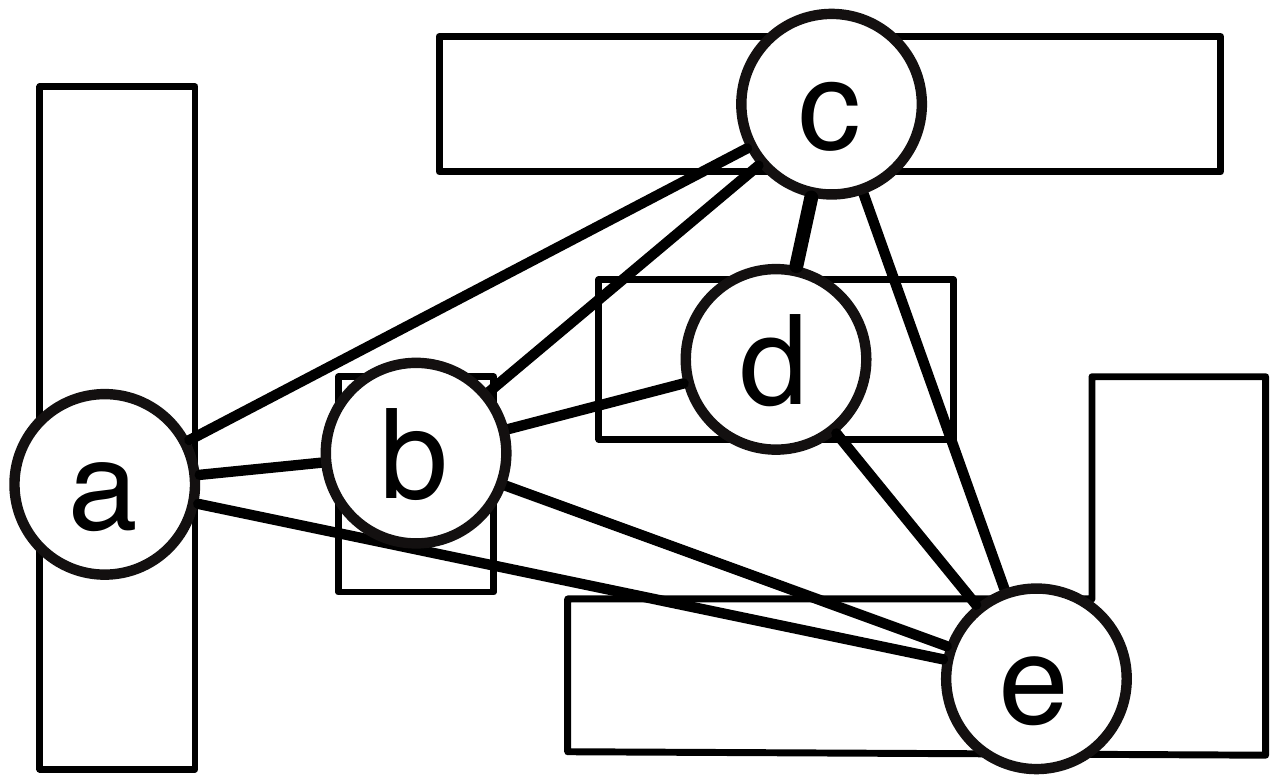}}
  \hspace{.2in}
  \subfigure[]{\includegraphics[width=0.14\textwidth]{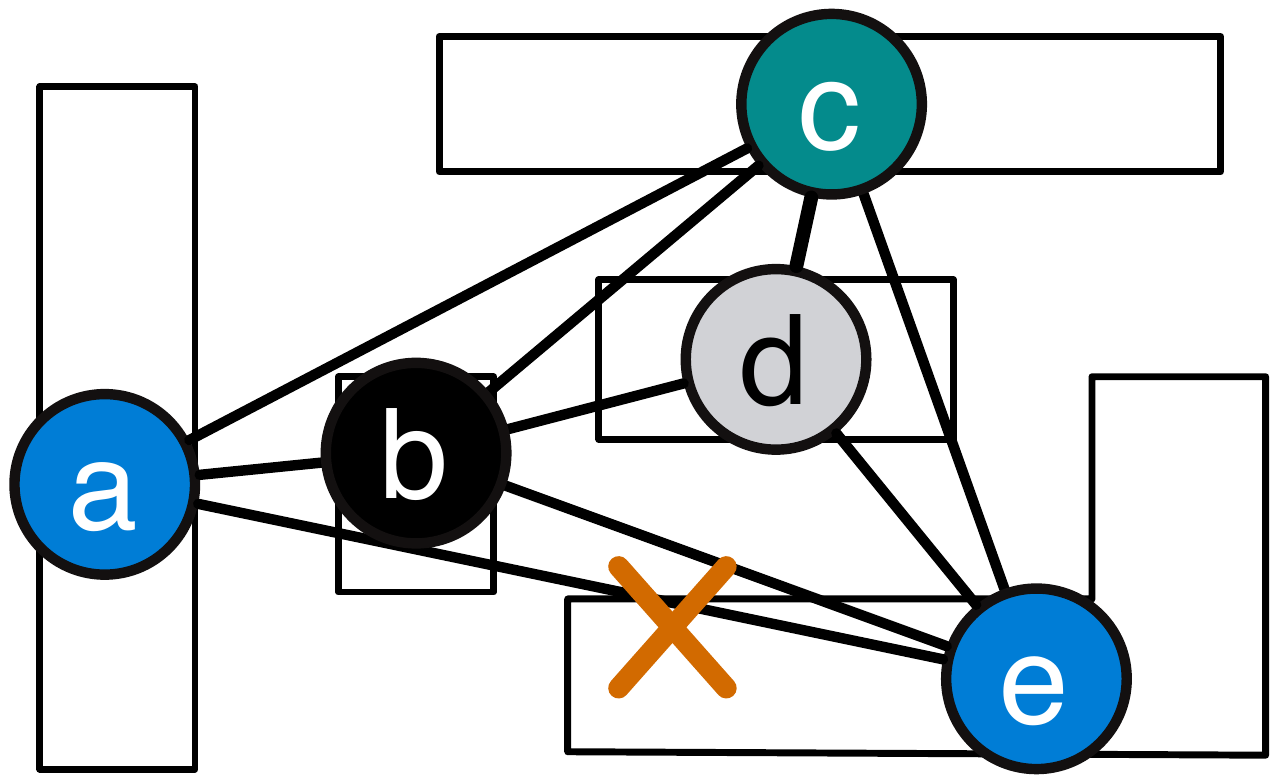}}
  \hspace{.2in}
  \subfigure[]{\includegraphics[width=0.14\textwidth]{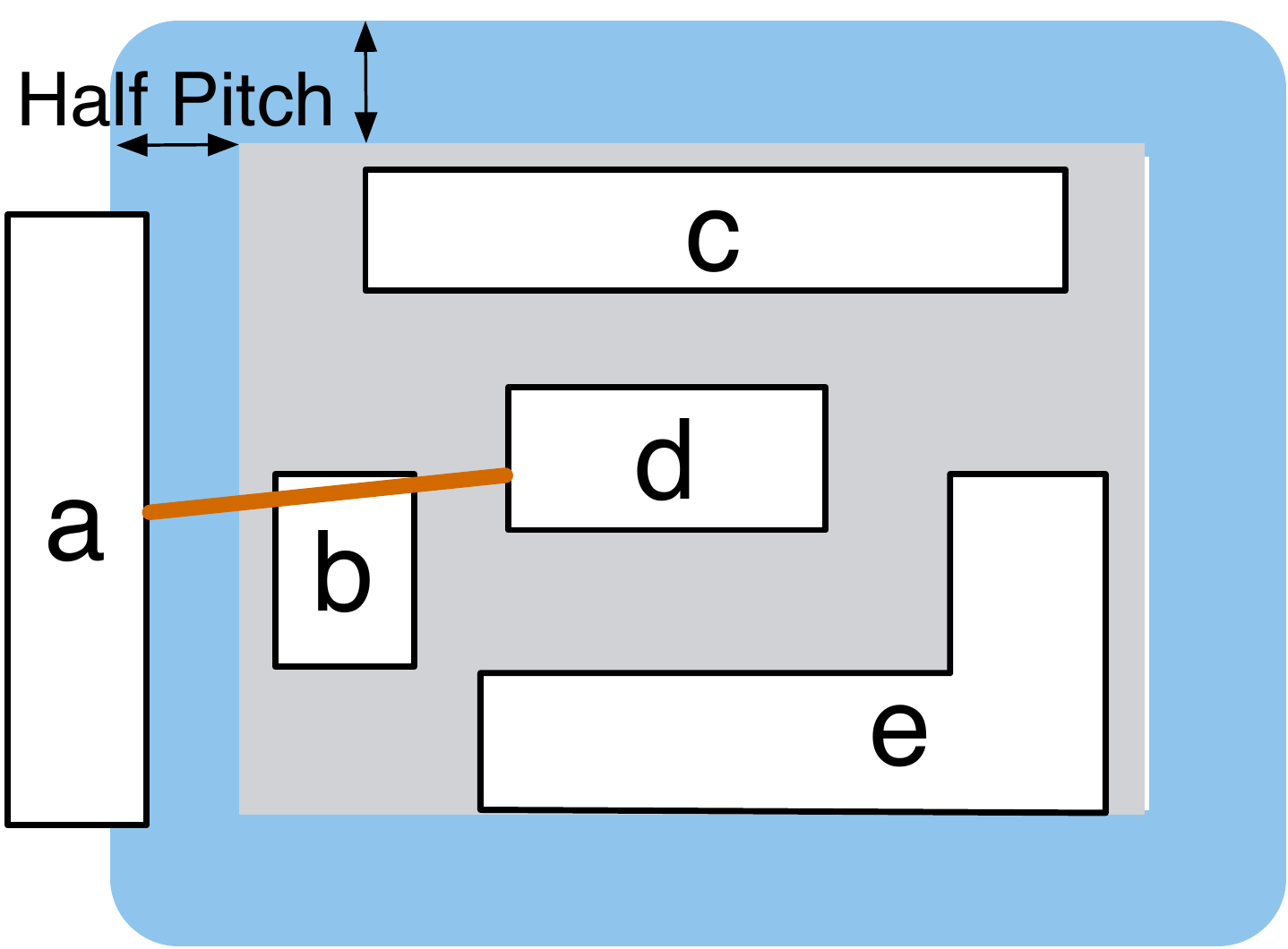}}
  \hspace{.2in}
  \subfigure[]{\includegraphics[width=0.14\textwidth]{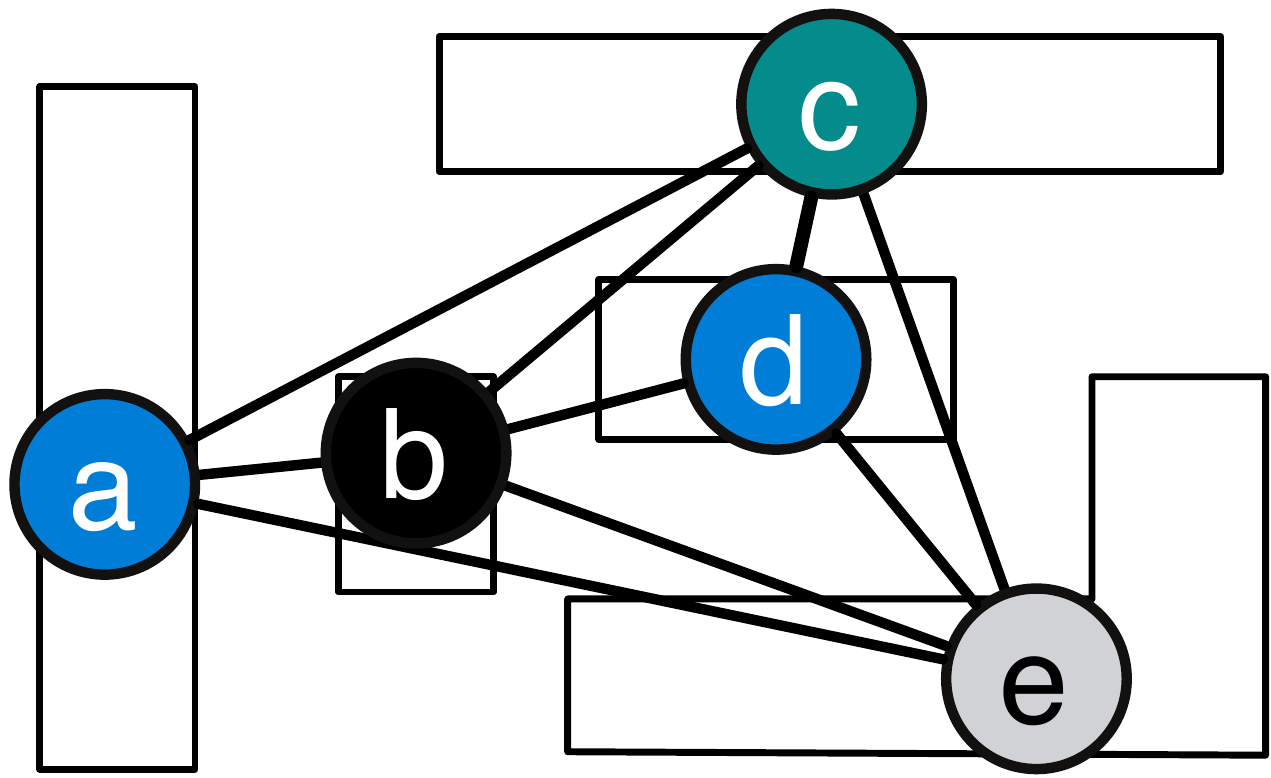}}
  \caption{
    (a) Decomposition graph;
    (b) Greedy coloring with one conflict;
    (c) $a$ is detected as color-friendly to $d$;
    (d) Coloring considering color-friendly rules.
  }
  \label{fig:color_friendly}
  \vspace{-.1in}
\end{figure}

In the second stage (lines 5-9), all remaining vertices would be assigned colors one by one.
%, thus the runtime complexity is linear.
However, color assignment through one specific order may be stuck at \textit{local optimum} which stems from the greedy nature.
For example, given a decomposition graph in Fig. \ref{fig:color_friendly} (a),
if the coloring order is \textit{a-b-c-d-e}, when vertex $d$ is greedily selected grey color,
the following vertex $e$ cannot find any color without conflict (see Fig. \ref{fig:color_friendly} (b)).
In other words, vertex ordering significantly impacts the coloring result.

To alleviate the impact of vertex ordering, two strategies are proposed.
The first strategy is called \textbf{color-friendly rules}, as in Definition \ref{def:friend}.
In Fig. \ref{fig:color_friendly} (c), all conflict neighbors of pattern $d$ are labeled inside a grey box.
Since the distance between $a$ and $d$ is within the range of $(min_s, min_s+hp)$, $a$ is color-friendly to $d$.
Interestingly, we discover a rule that for a complex/dense layout, color-friendly patterns tend to be with the same color.
Based on these rules, during linear color assignment, to determine one vertex color, instead of just comparing its conflict/stitch neighbors,
the colors of its color-friendly vertices would also be considered.
Detecting color-friendly vertices is similar to the conflict neighbor detection,
thus it can be finished during decomposition graph construction without much additional efforts.

\begin{mydefinition}[Color-Friendly]
\label{def:friend}
A pattern $a$ is color-friendly to pattern $b$, iff their distance is larger than $min_s$, but smaller than $min_s + hp$.
Here $hp$ is the half pitch.
\end{mydefinition}

Our second strategy is called \textbf{peer selection}, where three different vertex orders would be processed simultaneously,
and the best one would be selected as the final coloring solution (lines 6-8).
Although color assignment is solved thrice, since for each order the coloring is in linear time,
the total computational time is still linear.
\iffalse
More discussion regarding the peer selection can be found in Appendix \ref{sec:app_linear}.
\fi

In the third stage (line 10), post-refinement greedily checks each vertex to see whether the solution can be further improved.

% ============================================================
%             Last: complexity analyzing
% ============================================================
%Now we analyze the time complexity of our approach.
For a decomposition graph with color-friendly information and $n$ vertices,
in the first stage vertex removal/pop up can be finished in $O(n)$.
In the second stage, as mentioned above the coloring needs $O(n)$.
In post-refinement stage, all vertices are traveled once, which requires $O(n)$ time.
Therefore, the total complexity is $O(n)$.

\section{Graph Division for QPLD}
\label{sec:graph}

Graph division is a technique that partitions the whole decomposition graph into a set of components,
then the color assignment on each component can be solved independently.
%It has been proven powerful in double/triple patterning layout decomposition.
In our framework, the techniques extended from previous work are summarized as follows,
(1) Independent Component Computation
\cite{DPL_ISPD09_Yuan,DPL_ISPD2010_Xu,DPL_ICCAD2011_Tang,TPL_ICCAD2011_Yu,TPL_DAC2012_Fang,TPL_ICCAD2012_Tian,TPL_DAC2013_Kuang,TPL_ICCAD2013_Yu}; 
%DPL_ASPDAC2010_Yang,
(2) Vertex with Degree Less than 3 Removal
\cite{TPL_ICCAD2011_Yu,TPL_DAC2012_Fang,TPL_DAC2013_Kuang,TPL_ICCAD2013_Yu}
\footnote{In QPLD problem, the vertices with degree less than 4 would be detected and removed temporally.};
%(3) 2-Edge-Connected Component Computation \cite{TPL_ICCAD2011_Yu,TPL_DAC2012_Fang,TPL_DAC2013_Kuang,TPL_ICCAD2013_Yu};
(3) 2-Vertex-Connected Component Computation \cite{TPL_DAC2012_Fang,TPL_DAC2013_Kuang,TPL_ICCAD2013_Yu}.
%In our framework we apply all these techniques, with a slight modification.
%In our framework we apply some techniques that have been proven powerful in double/triple patterning layout decomposition study,
%summarized as follows.

%Besides, we follow prior motivations while approaching a new graph division via a general all-pair min-cut structure of GH-tree \cite{1961GHTree,1990GHTree}.
%GH-tree based cut removal can provide not only better problem size reduction, but also good viability to be applied in general K-patterning problems.
%GH-tree based graph division has the following significant strength and advantages.
%(By Tsung-Wei: it's better to show some "strength" and "advantages" in the following items instead of just the title simply.) 

% ==========================================================================
%                 Subsection: GH-Tree based Division
% ==========================================================================
\subsection{GH-Tree based 3-Cut Removal}

Another technique, \textit{cut removal}, has been proven powerful in double/patterning layout decomposition
\cite{DPL_ICCAD2011_Tang,TPL_ICCAD2011_Yu,TPL_DAC2012_Fang}.
A cut of a graph is an edge whose removal disconnects the graph into two components.
The definition of cut can be extended to 2-cur (3-cut), which is a double (triplet) of edges whose removal would disconnect the graph.
However, different from the 1-cut and 2-cut detection that can be finished in linear time \cite{TPL_DAC2012_Fang},
3-cut detection is much more complicated.
%still little known in EDA community.
In this subsection we propose an effective 3-cut detection method.
Besides, our method can be easily extended to detect any K-cut (K $\ge$ 3).

\begin{figure}[htb]
  \centering
  \vspace{-.1in}
  \includegraphics[width=0.4\textwidth]{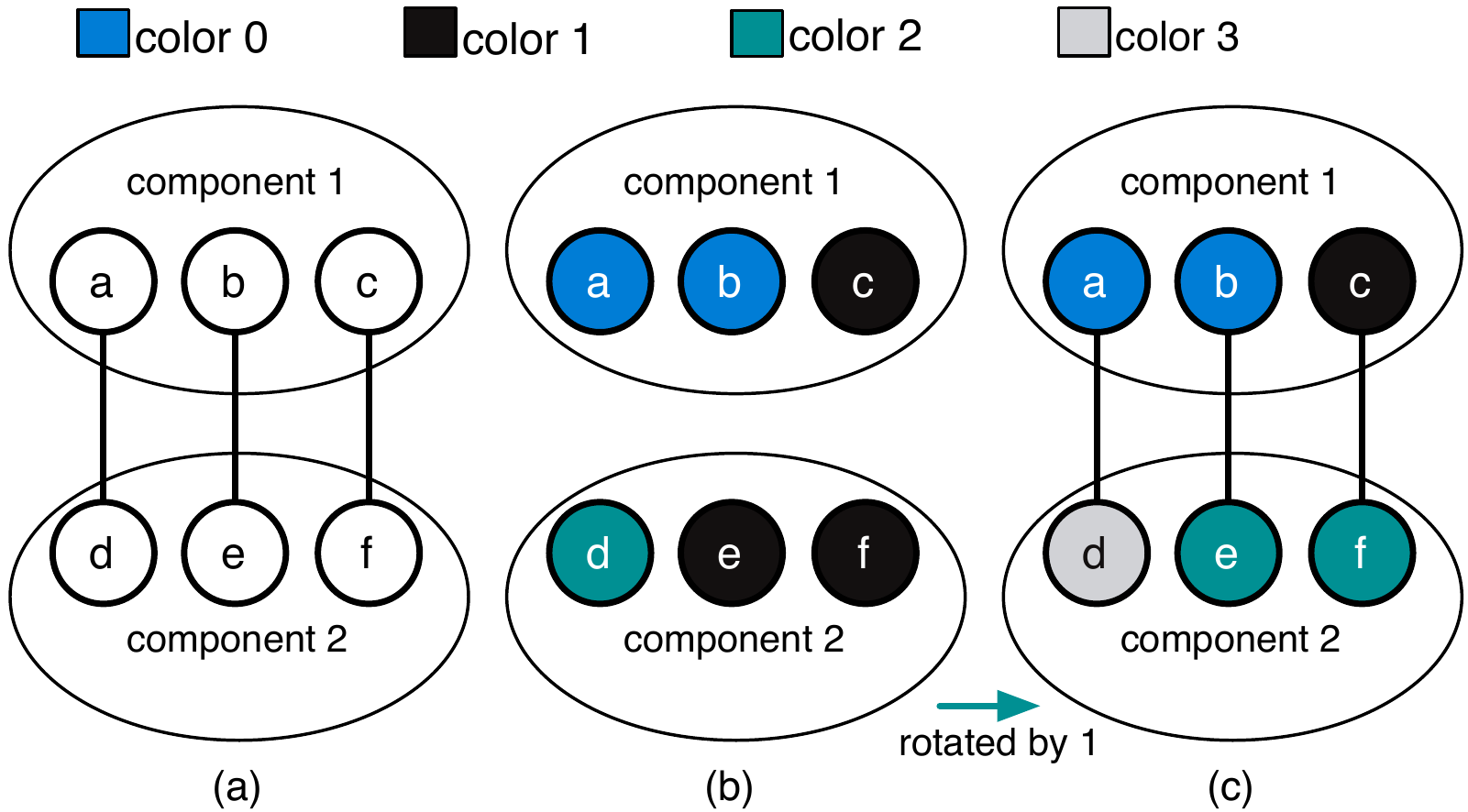}
  \caption{An example of 3-cut detection and removal.
  %~(a) A 3-cut between two components;
  %~(b) Color assignments on two components, separately;
  %~(c) Conflict removal through color rotation.
  }
  \label{fig:example_3cut}
\end{figure}

Fig . \ref{fig:example_3cut} (a) shows a graph with a 3-cut ($a-d, b-e, c-f$),
and two components can be derived by removing this 3-cut.
After color assignment on two components, for each cut edge, if the colors of the two endpoints are different,
the two components can be merged directly.
Otherwise, a \textit{color rotation} operation is required to one component.
For vertex $v$ in graph, we denote $c(v)$ as its color, where $c(v) \in \{0, 1, 2, 3\}$. 
Vertex $v$ is said to be rotated by $i$, if $c(v)$ is changed to $(c(v)+i)\%4$.
It is easy to see that all vertices in one component should be rotated by the same value, so no additional conflict is introduced within the component.
An example of such color rotation operation is illustrated in Fig. \ref{fig:example_3cut} (b)-(c),
where conflict between vertices $c, f$ would be removed to interconnect two components together. 
Here all the vertices in component 2 are rotated by 1 (see Fig. \ref{fig:example_3cut} (c)).
We have the following Lemma:

\begin{mylemma}
\label{lem:3cut}
In QPLD problem, color rotation after interconnecting 3-cut does not increase the conflict number.
\end{mylemma}

%We divide the decomposition graph through 3-cut into a set of components.
%By Lemma \ref{lem:3cut}, each component is independent of each other.
%To be more precise, the entire problem is partitioned into a set of independent subproblems which are relatively manageable and less computationally expensive. 
%Because of Lemma \ref{lem:3cut}, if we divide the decomposition graph through 3-cut,
%the whole problem can be partitioned into a set of independent sub-problems, so the layout decomposition framework can be speedup without introducing additional conflict.
%Here we show a novel method to compute all 3-cuts.
%Note that our method is generic that it can be extended to arbitrary $K$-cut.

In addition, to detect all 3-cuts, we have the following Lemma:

\begin{mylemma}
\label{lem:3cut2}
If the minimum cut between two vertices $v_i$ and $v_j$ is less than 4,
then $v_i, v_j$ belong to different components that divided by a 3-cut.
\end{mylemma}

%Due to space limit, the detailed proof is omitted.
Based on Lemma \ref{lem:3cut2}, we can see that if the cut between vertices $v_i, v_j$ is larger or equal to $4$ edges, $v_i, v_j$ should belong to the same component.
%One example of Lemma \ref{lem:3cut2} is shown in Fig. \ref{fig:GHTree} (a).
%The minimum cut between $a$ and $b$ is 3, therefore $a, b$ can belong to different components.
%In the graph the value on the edge represents its weight, the minimum cut between $a$ and $b$ is 2.
One straightforward 3-cut detection method is to compute the minimum cuts for all the $\{s-t\}$ pairs.
%then the division can be derived based on all the cut values.
However, for a decomposition graph with $n$ vertices, there are $n(n - 1)/2$ pairs of vertices.
Computing all these cut pairs may spend too long runtime, which is impractical for complex layout design.

\begin{figure}[tb]
  \centering
  \subfigure[] {\includegraphics[width=0.12\textwidth]{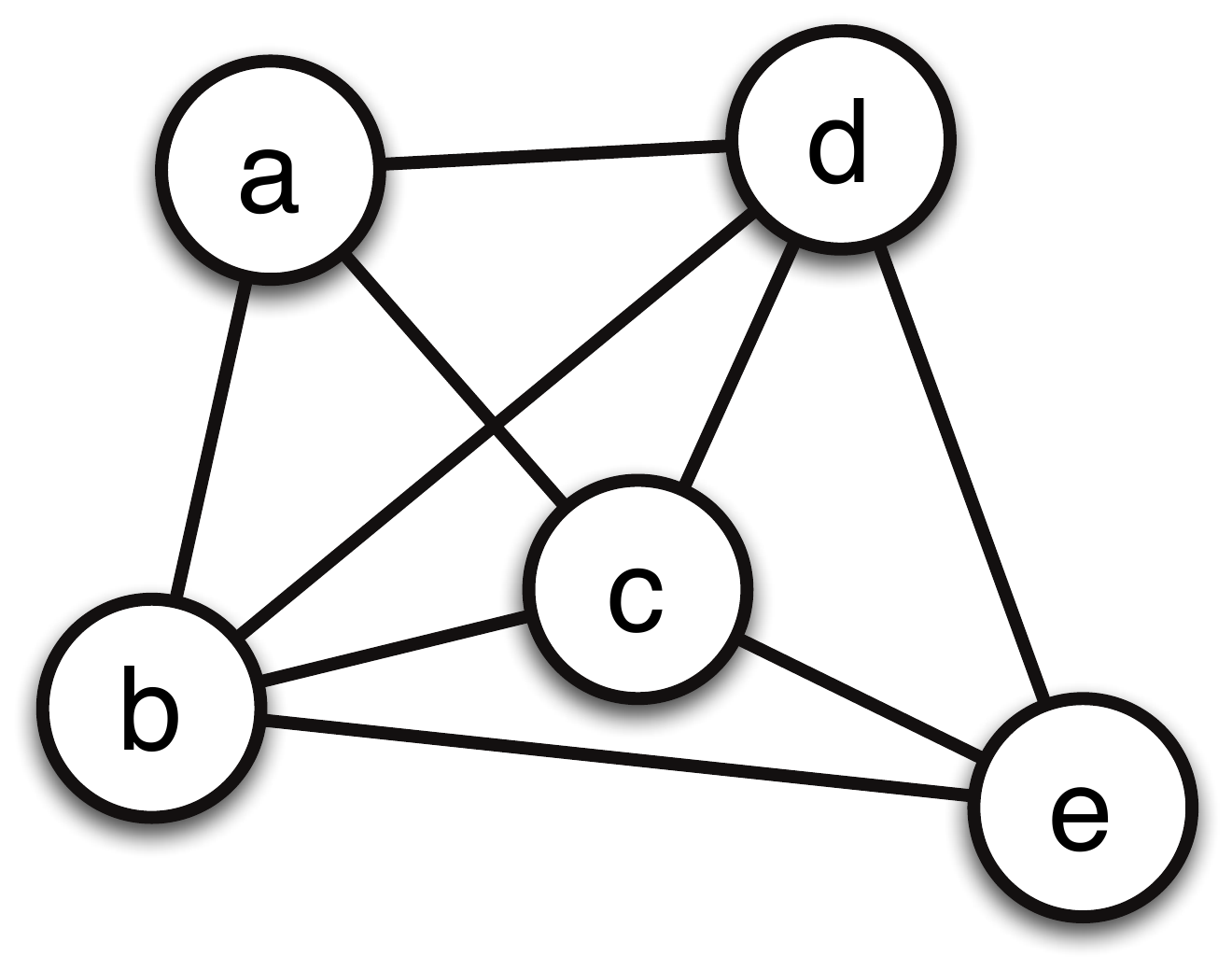}}
  \hspace{.1in}
  \subfigure[] {\includegraphics[width=0.10\textwidth]{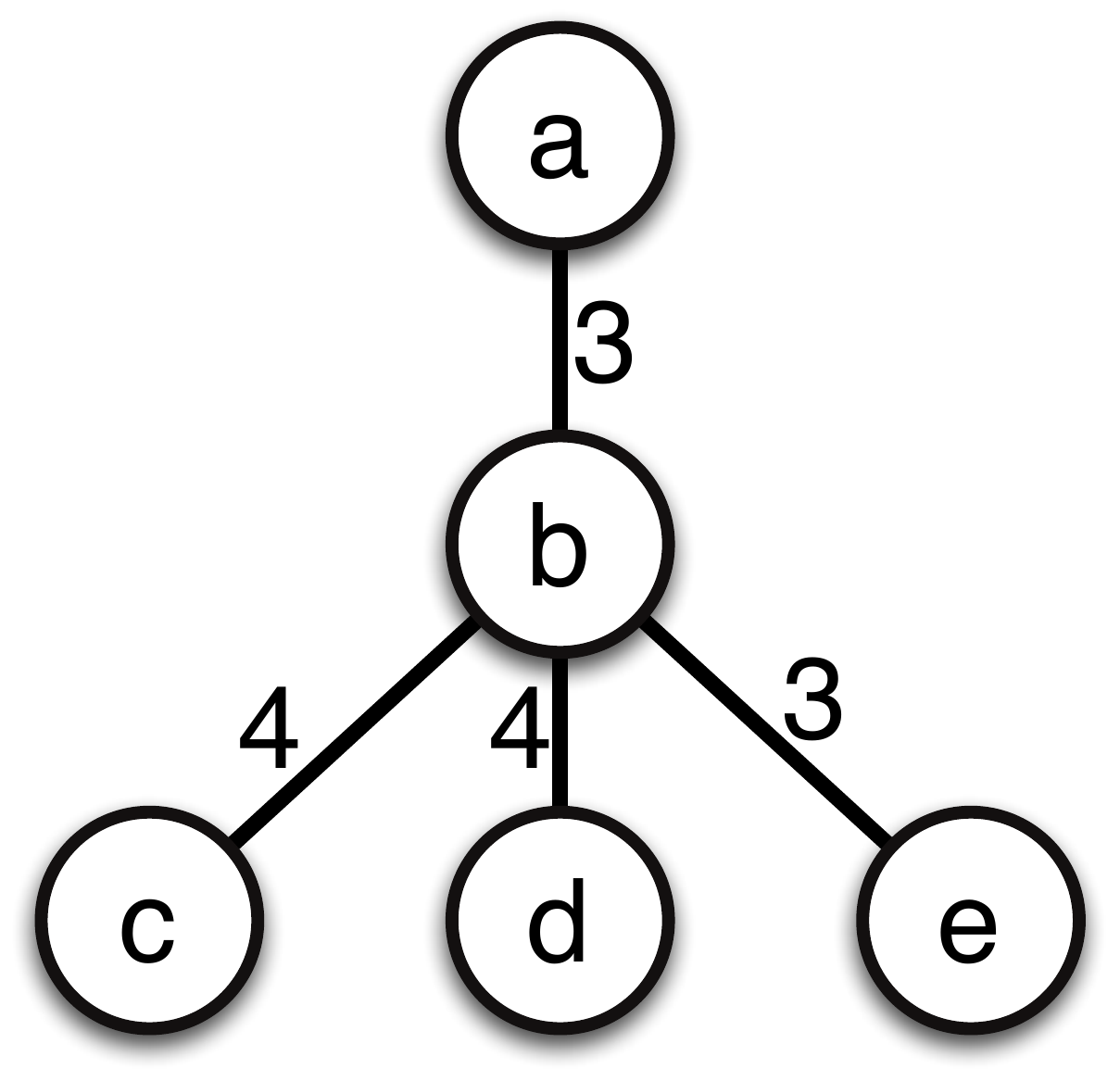}}
  \hspace{.1in}
  \subfigure[] {\includegraphics[width=0.10\textwidth]{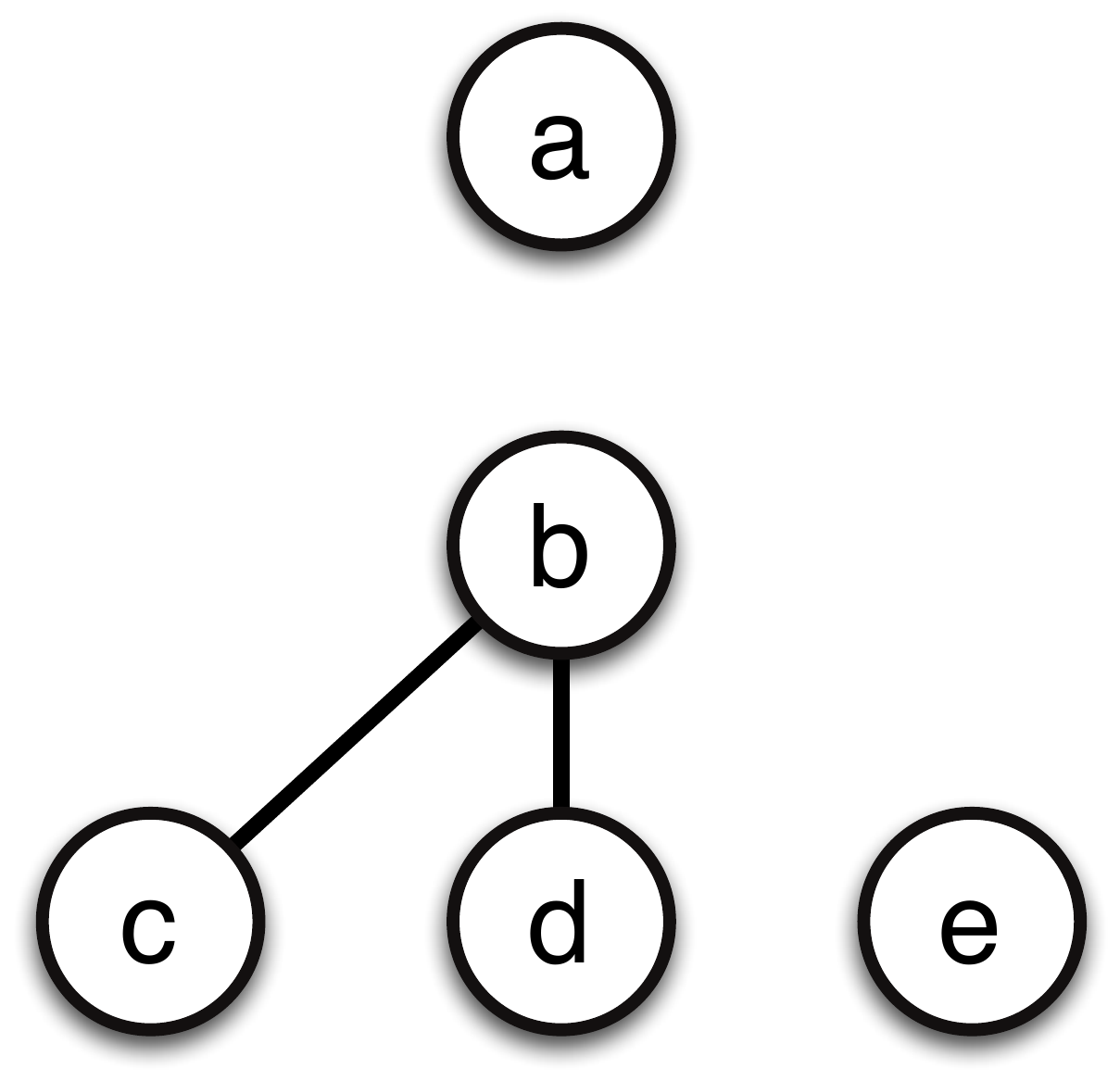}}
  %\hspace{.1in}
  %\subfigure[] {\includegraphics[width=0.10\textwidth]{GHTree3}}
  \caption{~(a) Decomposition graph;~(b) Corresponding GH-tree;~(c) Components after 3-cut removal.}
  \label{fig:GHTree}
  \vspace{-.1in}
\end{figure}

Gomory and Hu \cite{1961GHTree} showed that the cut values between all the pairs of vertices can be computed by solving only $n-1$ network flow problems on graph $G$.
Furthermore, they showed that the flow values can be represented by a weighted tree $T$ on the $n$ vertices,
where for any pair of vertices $(v_i, v_j)$, if $e$ is the minimum weight edge on the path from $v_i$ to $v_j$ in $T$,
then the minimum cut value from $v_i$ to $v_j$ in $G$ is exactly the weight of $e$.
Such a weighted tree $T$ is called Gomory-Hu tree (GH-tree).
%\cite{1961GHTree} is a weighted tree that represents the minimum $s-t$ cuts for all $s-t$ pairs in an undirected graph.
%GH tree \cite{1961GHTree} is used to search the cut for each pair.
%The GH-tree can be constructed in $|V|-1$ maximum flow (minimum cut) computations.
%We first provide a precise definition of a GH tree and some preliminary,
%then we discuss how GH tree can be applied to achieve graph simplification.
For example, given the decomposition graph in Fig. \ref{fig:GHTree} (a), the corresponding GH-tree is shown in Fig. \ref{fig:GHTree} (b),
where the value on edge $e_{ij}$ is the minimum cut number between vertices $v_i$ and $v_j$.
Because of Lemma \ref{lem:3cut2}, to divide the graph through 3-cut removal, all the edges with value less than 4 would be removed.
The final three components are in Fig. \ref{fig:GHTree} (c).
%In Fig. \ref{fig:GHTree} (b), we illustrate a GH-tree obtained from the given decomposition graph in Fig. \ref{fig:GHTree} (a).

%\begin{mydefinition}[GH Tree]
%Given a network $G=(V, E)$, a GH tree $T = (V, F)$ obtained from $G$ is a weighted tree with the same set of vertices $V$ with the two following properties:
%(1) Equivalent flow tree; (2) Cut property.
%\end{mydefinition}

\begin{algorithm}[htb]
\caption{GH-tree based 3-Cut Removal}
\label{alg:ghtree}
\begin{algorithmic}[1]
  \Require Decomposition graph $G=\{V, CE, SE\}$;
  \State Construct GH-tree as in \cite{1990GHTree};
  \State Remove the edges with weight $<$ 4;
  \State Compute connected components on remaining GH-tree;
  \For {each component}
    \State Color assignment on this component;
  \EndFor
  %\State Color assignment on each component independently;
  \State Color rotation to interconnect all components;
\end{algorithmic}
\end{algorithm}

The procedure of the 3-cut removal is shown in Algorithm \ref{alg:ghtree}.
Firstly we construct GH-tree based on the algorithm by \cite{1990GHTree} (line $1$).
Dinic's blocking flow algorithm \cite{FLOW_1970_Dinic} is applied to help GH-tree construction.
%In order to calculate the GH-tree, several maximum flow problems need to be solved.
%Here we apply \textit{Dinic} algorithm \cite{FLOW_1970_Dinic} to solve the maximum flow problem.
Then all edges in the GH-tree with weights less than four are removed (line $2$).
After solving the connected component problem (line $3$), we can assign colors to each component separately (lines $4-5$).
At last color rotation is applied to interconnect all 3-cuts back (line $6$).

\section{General K-Patterning Layout Decomposition}
\label{sec:extend}

In this section, we demonstrate that our layout decomposition framework is generalizable to K-patterning layout decomposition, where $K > 4$.

\begin{mytheorem}
SDP formulation in (\ref{eq:mpl_sdp}) can provide $v_i \cdot v_j$ pairs for K-patterning color assignment problem.
\label{thm:ksdp}
\end{mytheorem}

\begin{figure}[h]
\vspace{-.2in}
\begin{align}
  \label{eq:mpl_sdp}
  \textrm{min}  & \sum_{e_{ij} \in CE} ( \vec{v_i} \cdot \vec{v_j}+\frac{1}{k-1}) + \alpha \sum_{e_{ij} \in SE} ( 1 - \vec{v_i} \cdot \vec{v_j} )\\
  \textrm{s.t}.\ \	
    & \vec{v_i} \cdot \vec{v_i} = 1 ,               \ \ \ \forall i \in V			        \notag\\
    & \vec{v_i} \cdot \vec{v_j} \ge -\frac{1}{k-1},   \ \ \ \forall e_{ij} \in CE      \notag
\end{align}
\vspace{-.2in}
\end{figure}

%Due to space limit, the detailed proof is omitted.
We can see that if $K=4$, formulation (\ref{eq:mpl_sdp}) equivalents to (\ref{eq:sdp}).
Rephrasing both the SDP formulation in (\ref{eq:mpl_sdp}) and backtrack method in Algorithm \ref{alg:backtrack},
the color assignment problem for K-patterning can be resolved.
In addition, the linear color assignment algorithm in Section \ref{sec:linear} can be extended to general K-patterning problem as well.

All the graph division techniques in Section \ref{sec:graph} can be extended here.
Besides, we draw the following Theorem:

\begin{mytheorem}
For K-patterning layout decomposition problem, dividing graph through ($K-1$)-cut does not increase the final conflict number.
\label{thm:graph}
\end{mytheorem}

The proof can be provided by extending Lemma \ref{lem:3cut}. % and is omitted due to page limitation.
%Due to space limit, the detailed proof is omitted.
%But it is easy to see that the proof of Theorem \ref{theo:3cut} can be extended here.
Based on Theorem \ref{thm:graph}, GH-tree based cut removal in Section \ref{sec:graph} can be applied here to search all $(K-1)$-cuts.
%It shall be noted that after constructing GH-tree, all the edges with weight less than $K$ would be removed.
That is, after constructing GH-tree, all edges with weight less than $K$ are removed.

% table 1: comparison on QPLD
\begin{table*}[tb]
\centering
%\footnotesize
\scriptsize
\caption{Comparison for Quadruple Patterning}
\label{tab:qpld}
%{{{
\begin{tabular}{|c|ccc|ccc|ccc|ccc|}
  \hline \hline
  \multirow{2}{*}{Circuit}&\multicolumn{3}{c|}{ILP}             &\multicolumn{3}{c|}{SDP+Backtrack}   
                          &\multicolumn{3}{c|}{SDP+Greedy}      &\multicolumn{3}{c|}{Linear}\\
  \cline{2-13}&cn\#  &st\#   & CPU(s)         &cn\#  &st\#    &CPU(s)       &cn\#  &st\#    &CPU(s)       &cn\#  &st\#  &CPU(s)\\
  \hline                                                                    
  C432        &2     &0      &0.6             &2     &0       &0.24        &2     &0      &0.02        &2     &1       &0.001   \\
  C499        &1     &4      &0.7             &1     &4       &0.16        &1     &4      &0.05        &1     &4       &0.001   \\
  C880        &1     &0      &0.3             &1     &0       &0.02        &1     &0      &0.02        &1     &2       &0.001   \\
  C1355       &0     &4      &0.6             &0     &4       &0.1         &0     &4      &0.04        &0     &4       &0.001   \\
  C1908       &2     &3      &1.0             &2     &3       &0.28        &2     &3      &0.09        &2     &4       &0.001   \\
  C2670       &0     &6      &1.1             &0     &6       &0.16        &0     &6      &0.1         &0     &7       &0.001   \\
  C3540       &1     &3      &1.1             &1     &3       &0.09        &2     &2      &0.05        &1     &3       &0.001   \\
  C5315       &1     &13     &2.8             &1     &13      &0.6         &2     &12     &0.24        &1     &15      &0.002   \\
  C6288       &9     &0      &2.3             &9     &0       &0.36        &9     &0      &0.17        &9     &1       &0.001   \\
  C7552       &2     &13     &3.4             &2     &13      &0.6         &3     &12     &0.22        &2     &18      &0.003   \\
  S1488       &0     &6      &0.7             &0     &6       &0.05        &4     &2      &0.01        &0     &6       &0.001   \\
  S38417      &20    &549    &1226.7          &20    &551     &6.6         &142   &429    &2.7         &21    &576     &0.03    \\
  S35932      &N/A   &N/A    &$>$3600         &50    &1745    &28.7        &460   &1338   &16.4        &64    &1927    &0.15    \\
  S38584      &N/A   &N/A    &$>$3600         &41    &1653    &21.1        &470   &1224   &10.4        &47    &1744    &0.12    \\
  S15850      &N/A   &N/A    &$>$3600         &42    &1462    &18          &420   &1084   &7.8         &48    &1571    &0.11    \\
  \hline                                            
  avg.        &-     &-      &$>$802.7        &11.5  &364.0   &5.14         &101.2 &274.7   &2.56       &13.3&392.2    &0.03 \\
  ratio       &-     &-      &\textbf{$>$156.3}                 &\textbf{1.0}   &\textbf{1.0}  &\textbf{1.0}
              &\textbf{8.83} &\textbf{0.75}  &\textbf{0.49}    &\textbf{1.15}  &\textbf{1.08} &\textbf{0.005}\\
  \hline \hline %\toprule
\end{tabular}
\vspace{-.1in}
%}}}
\end{table*}

\section{Experimental Results}
\label{sec:result}

\begin{figure}[htb]
  \centering
  \vspace{-.1in}
  \subfigure[]{\includegraphics[width=0.14\textwidth]{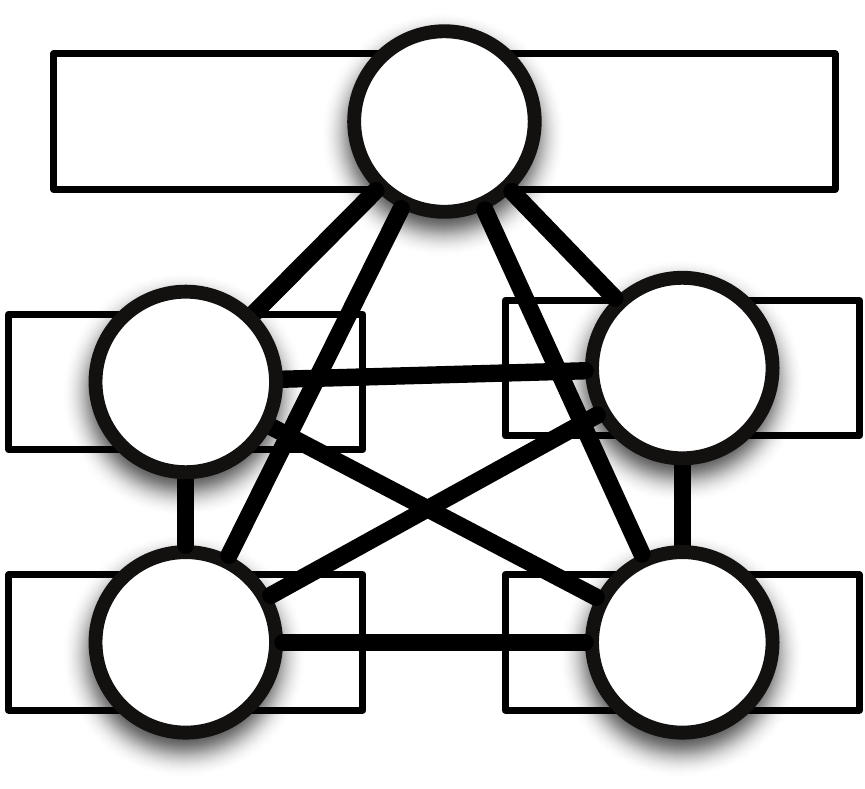}}
  \hspace{.2in}
  \subfigure[]{\includegraphics[width=0.20\textwidth]{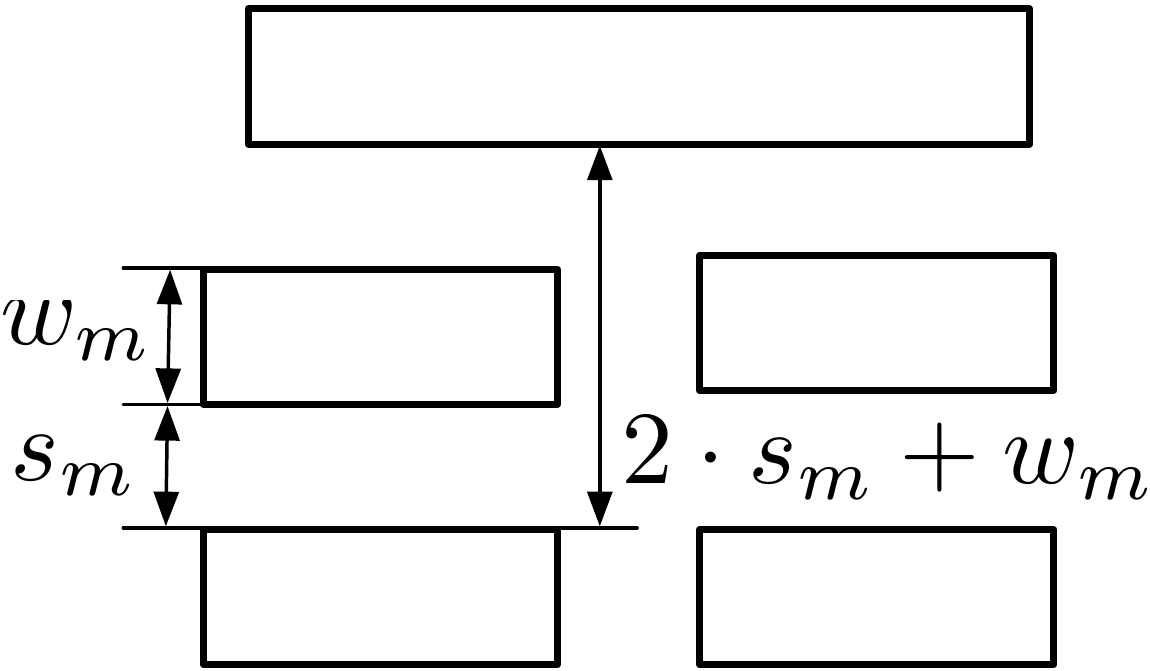}}
  \caption{$min_s = 2 \cdot s_m + w_m$ may cause $K_5$ structure.}
  \label{fig:MPL_QPL_mins}
  \vspace{-.2in}
\end{figure}

We implemented the proposed layout decomposition algorithms in C++, and tested on a Linux machine with 2.9GHz CPU.
We choose GUROBI \cite{Gurobi} as the integer linear programming (ILP) solver, and CSDP \cite{CSDP} as the SDP solver.
The benchmarks in \cite{TPL_ICCAD2011_Yu,TPL_DAC2012_Fang} are used as our test cases.
We scale down the Metal1 layer to 20nm half pitch.
Both the minimum feature width $w_m$ and the minimum spacing between features $s_m$ are 20nm.
From Fig. \ref{fig:MPL_QPL_mins} we can see that when minimum coloring distance $min_s = 2 \cdot s_m + w_m = 60$nm,
even one dimension regular patterns can be a $K_5$ structure,
which is not 4-colorable or planar \cite{1930Kuratowski}.
In our experiments, for quadruple patterning $min_s$ is set as $2 \cdot s_m + 2 \cdot w_m = 80$nm,
while for pentuple patterning $min_s$ is set as $3 \cdot s_m + 2.5 \cdot w_m = 110$nm.
When larger $min_s$ is applied, there are too many native conflicts in layouts,
as the benchmarks are not multiple patterning friendly.

\subsection{Quadruple Patterning}

First we compare different color assignment algorithms for quadruple patterning, and the results are listed in Table \ref{tab:qpld}.
``\textbf{ILP}'', ``\textbf{SDP+Backtrack}'', ``\textbf{SDP+Greedy}'' and ``\textbf{Linear}'' denote ILP formulation,
SDP followed by backtrack mapping (Section \ref{sec:sdp}),
SDP followed by greedy mapping,
and linear color assignment (Section \ref{sec:linear}), respectively.
Here we implement an ILP formulation extended from the triple patterning work \cite{TPL_ICCAD2011_Yu}.
\iffalse
(details in Appendix \ref{sec:app_ilp})
\fi
In SDP+Greedy, a greedy mapping from \cite{TPL_ICCAD2011_Yu} is applied.
All the graph division techniques, including GH-tree based division, are applied.
The columns ``cn\#" and ``st\#" denote the conflict number and the stitch number, respectively.
%Column ``cost'' is the cost function, which is set as $\textrm{conflict\#} + 0.1 \times \textrm{stitch\#}$.
Column ``CPU(s)" is color assignment time in seconds.
%In column ``OPT'', ``$\checkmark$'' means the solution is optimal.

From Table \ref{tab:qpld} we can see that for small cases the ILP formulation can achieve best performance in terms of conflict number and stitch number.
However, for large cases (S38417, S35932, S38584, S15850) ILP suffers from long runtime problem that none of them can be finished in one hour.
Compared with ILP, SDP+Backtrack can achieve near-optimal solutions, i.e.,
in every case the conflict number is optimal, while only in one case 2 more stitches are introduced.
SDP+Greedy method can achieve 2$\times$ speedup against SDP+Backtrack.
But the performance of SDP+Greedy is not good that for complex designs hundreds of additional conflicts are reported.
The linear color assignment can achieve around 200$\times$ speedup against SDP+Backtrack, while only 15\% more conflicts and 8\% more stitches are reported.

% table 3: k=5, mindp=220
\begin{table*}[tb]
\scriptsize
\centering
\caption{Comparison for Pentuple Patterning}
\label{tab:kmask}
%{{{
\begin{tabular}{|c|ccc|ccc|ccc|}
  \hline \hline
  \multirow{2}{*}{Circuit}  &\multicolumn{3}{c|}{SDP+Backtrack}     &\multicolumn{3}{c|}{SDP+Greedy}   &\multicolumn{3}{c|}{Linear}\\
  \cline{2-10}              &cn\#  &st\#  &CPU(s)         &cn\#  &st\#  &CPU(s)                 &cn\#  &st\#  &CPU(s)\\
  \hline                                                                                    
  C6288      &19  &2   &2.4         &19  &2   &0.49         &19  &5   &0.005  \\
  C7552      &1   &1   &0.3         &1   &1   &0.05         &1   &4   &0.001  \\
  S38417     &0   &4   &1.45        &0   &4   &0.21         &0   &4   &0.001  \\
  S35932     &5   &20  &8.11        &5   &20  &0.62         &5   &25  &0.009  \\
  S38584     &3   &4   &1.66        &7   &3   &0.3          &3   &6   &0.008  \\
  S15850     &6   &5   &2.7         &7   &5   &0.4          &5   &15  &0.007  \\
  \hline                            
  avg.       &5.7 &6.0 &2.77        &6.5 &5.83&0.35         &5.5 &9.8 &0.005 \\
  ratio      &\textbf{1.0}  &\textbf{1.0}   &\textbf{1.0}
             &\textbf{1.15} &\textbf{0.97}  &\textbf{0.12}
             &\textbf{0.97} &\textbf{1.64}  &\textbf{0.002} \\
  \hline \hline
\end{tabular}
\vspace{-.1in}
%}}}
\end{table*}

% ============== 5 Patterning Layout Decomposition =============
\subsection{Pentuple Patterning}

We further compare the algorithms for pentuple patterning, that is, $K=5$.
To our best knowledge there is no exact ILP formulation for pentuple patterning in literature.
Therefore we consider three baselines, i.e., SDP+Backtrack, SDP+Greedy, and Linear.
All the graph division techniques are applied.
Table \ref{tab:kmask} evaluates six most dense cases.
We can see that compared with SDP+Backtrack, SDP+Greedy can achieve around 8$\times$ speedup, but 15\% more conflicts are reported.
In terms of runtime, linear color assignment can achieve 500$\times$ and 60$\times$ speedup, against SDP+Backtrack and SDP+Greedy, respectively.
In terms of performance, linear color assignment reports the best conflict number minimization, but more stitches may be introduced.

Interestingly, we observe that when a layout is multiple patterning friendly,
color-friendly rules can provide a good guideline, thus linear color assignment can achieve high performance in terms of conflict number.
However, when a layout is very complex or involving many native conflicts,
linear color assignment reports more conflicts than SDP+Backtrack.
%SDP+Backtrack reports less conflicts.
One possible reason is that the color-friendly rules are not good in modeling global conflict minimization,
but both SDP and backtrack provide a global view.

\vspace{-.1in}
\section{Conclusions}
\label{sec:conclu}

% ======= by David
In this paper we have proposed the first layout decomposition framework for quadruple patterning and beyond.
Experimental evaluations have demonstrated that our algorithm is effective and efficient to obtain high quality solution.
As continuing scaling of technology node to sub-10nm, MPL may be a promising manufacturing solution.
%A dedicated layout decomposition tool is necessary to assist in the whole process.
We believe this paper will stimulate more future research into this field, thereby facilitating the advancement of MPL technology.

% ======= by Tsung-Wei
\iffalse
In this paper we have proposed the first layout decomposition algorithm for quadruple patterning.
Experimental evaluations have demonstrated that our algorithm is effective and efficient to obtain high solution quality.
Moreover, the proposed algorithm is generalizable to multiple patterning lithography (MPL) which provides a bridge in conjunction with the entire problem family.
As continuing growth of technology node to sub-11nm, QPL/MPL turns out to be a promising manufacturing solution.
%A dedicated layout decomposition tool is necessary to assist in the whole process.
We believe this paper will stimulate more future research into this field, thereby facilitating the advance of MPL technology.
\fi

\vspace{-.1in}
\section*{Acknowledgment}

The authors would like to thank Tsung-Wei Huang for helpful discussions.
This work is supported in part by NSF, NSFC, SRC, Oracle, and Toshiba.

{
\vspace{-.1in}
\scriptsize
%\footnotesize
\bibliographystyle{IEEEtran}
\bibliography{./Ref/Bei,./Ref/Algorithm,./Ref/MPL,./Ref/Lith,./Ref/EUV,./Ref/EBL,./Ref/Partition}

% Generated by IEEEtran.bst, version: 1.13 (2008/09/30)
\begin{thebibliography}{10}
\providecommand{\url}[1]{#1}
\csname url@samestyle\endcsname
\providecommand{\newblock}{\relax}
\providecommand{\bibinfo}[2]{#2}
\providecommand{\BIBentrySTDinterwordspacing}{\spaceskip=0pt\relax}
\providecommand{\BIBentryALTinterwordstretchfactor}{4}
\providecommand{\BIBentryALTinterwordspacing}{\spaceskip=\fontdimen2\font plus
\BIBentryALTinterwordstretchfactor\fontdimen3\font minus
  \fontdimen4\font\relax}
\providecommand{\BIBforeignlanguage}[2]{{%
\expandafter\ifx\csname l@#1\endcsname\relax
\typeout{** WARNING: IEEEtran.bst: No hyphenation pattern has been}%
\typeout{** loaded for the language `#1'. Using the pattern for}%
\typeout{** the default language instead.}%
\else
\language=\csname l@#1\endcsname
\fi
#2}}
\providecommand{\BIBdecl}{\relax}
\BIBdecl

\vspace{.1in}

\bibitem{LITH_TCAD2013_Pan}
D.~Z. Pan, B.~Yu, and J.-R. Gao, ``Design for manufacturing with emerging
  nanolithography,'' \emph{IEEE Transactions on Computer-Aided Design of
  Integrated Circuits and Systems (TCAD)}, vol.~32, no.~10, pp. 1453--1472,
  2013.

\bibitem{LITH_ICCAD2012_Yu}
B.~Yu, J.-R. Gao, D.~Ding, Y.~Ban, J.-S. Yang, K.~Yuan, M.~Cho, and D.~Z. Pan,
  ``Dealing with {IC} manufacturability in extreme scaling,'' in \emph{Proc.
  ICCAD}, 2012, pp. 240--242.

\bibitem{DPL_ICCAD08_Kahng}
A.~B. Kahng, C.-H. Park, X.~Xu, and H.~Yao, ``Layout decomposition for double
  patterning lithography,'' in \emph{Proc. ICCAD}, 2008, pp. 465--472.

\bibitem{TPL_ICCAD2011_Yu}
B.~Yu, K.~Yuan, B.~Zhang, D.~Ding, and D.~Z. Pan, ``Layout decomposition for
  triple patterning lithography,'' in \emph{Proc. ICCAD}, 2011, pp. 1--8.

\bibitem{DPL_ISPD09_Yuan}
K.~Yuan, J.-S. Yang, and D.~Z. Pan, ``Double patterning layout decomposition
  for simultaneous conflict and stitch minimization,'' in \emph{Proc. ISPD},
  2009, pp. 107--114.

\bibitem{DPL_ISPD2010_Xu}
Y.~Xu and C.~Chu, ``A matching based decomposer for double patterning
  lithography,'' in \emph{Proc. ISPD}, 2010, pp. 121--126.

\bibitem{DPL_ICCAD2011_Tang}
X.~Tang and M.~Cho, ``Optimal layout decomposition for double patterning
  technology,'' in \emph{Proc. ICCAD}, 2011, pp. 9--13.

\bibitem{TPL_DAC2012_Fang}
S.-Y. Fang, W.-Y. Chen, and Y.-W. Chang, ``A novel layout decomposition
  algorithm for triple patterning lithography,'' in \emph{Proc. DAC}, 2012, pp.
  1185--1190.

\bibitem{TPL_DAC2013_Kuang}
J.~Kuang and E.~F. Young, ``An efficient layout decomposition approach for
  triple patterning lithography,'' in \emph{Proc. DAC}, 2013, pp. 69:1--69:6.

\bibitem{TPL_ICCAD2013_Yu}
B.~Yu, Y.-H. Lin, G.~Luk-Pat, D.~Ding, K.~Lucas, and D.~Z. Pan, ``A
  high-performance triple patterning layout decomposer with balanced density,''
  in \emph{Proc. ICCAD}, 2013, pp. 163--169.

\bibitem{TPL_ICCAD2013_Zhang}
Y.~Zhang, W.-S. Luk, H.~Zhou, C.~Yan, and X.~Zeng, ``Layout decomposition with
  pairwise coloring for multiple patterning lithography,'' in \emph{Proc.
  ICCAD}, 2013, pp. 170--177.

\bibitem{TPLEC_SPIE2013_Yu}
B.~Yu, J.-R. Gao, and D.~Z. Pan, ``Triple patterning lithography ({TPL}) layout
  decomposition using end-cutting,'' in \emph{Proc. of SPIE}, vol. 8684, 2013.

\bibitem{TPL_ICCAD2012_Tian}
H.~Tian, H.~Zhang, Q.~Ma, Z.~Xiao, and M.~Wong, ``A polynomial time triple
  patterning algorithm for cell based row-structure layout,'' in \emph{Proc.
  ICCAD}, 2012, pp. 57--64.

\bibitem{DFM_ICCAD2013_Yu}
B.~Yu, X.~Xu, J.-R. Gao, and D.~Z. Pan, ``Methodology for standard cell
  compliance and detailed placement for triple patterning lithography,'' in
  \emph{Proc. ICCAD}, 2013, pp. 349--356.

\bibitem{TPL_ICCAD2013_Tian}
H.~Tian, Y.~Du, H.~Zhang, Z.~Xiao, and M.~Wong, ``Constrained pattern
  assignment for standard cell based triple patterning lithography,'' in
  \emph{Proc. ICCAD}, 2013, pp. 57--64.

\bibitem{1998Karger}
D.~Karger, R.~Motwani, and M.~Sudan, ``Approximate graph coloring by
  semidefinite programming,'' \emph{J. ACM}, vol.~45, pp. 246--265, March 1998.

\bibitem{1977Appel}
K.~Appel and W.~Haken, ``Every planar map is four colorable. part i:
  Discharging,'' \emph{Illinois Journal of Mathematics}, vol.~21, no.~3, pp.
  429--490, 1977.

\bibitem{1930Kuratowski}
C.~Kuratowski, ``Sur le probleme des courbes gauches en topologie,''
  \emph{Fundamenta mathematicae}, vol.~15, no.~1, pp. 271--283, 1930.

\bibitem{1996Robertson}
N.~Robertson, D.~P. Sanders, P.~Seymour, and R.~Thomas, ``Efficiently
  four-coloring planar graphs,'' in \emph{ACM Symposium on Theory of
  computing}, 1996, pp. 571--575.

\bibitem{1961GHTree}
R.~E. Gomory and T.~C. Hu, ``Multi-terminal network flows,'' \emph{Journal of
  the Society for Industrial \& Applied Mathematics}, vol.~9, no.~4, pp.
  551--570, 1961.

\bibitem{1990GHTree}
D.~Gusfield, ``Very simple methods for all pairs network flow analysis,''
  \emph{SIAM Journal on Computing}, vol.~19, no.~1, pp. 143--155, 1990.

\bibitem{FLOW_1970_Dinic}
E.~A. Dinic, ``Algorithm for solution of a problem of maximum flow in networks
  with power estimation,'' in \emph{Soviet Math. Dokl}, vol.~11, no.~5, 1970,
  pp. 1277--1280.

\bibitem{Gurobi}
``{GUROBI},'' \url{http://www.gurobi.com/html/academic.html}.

\bibitem{CSDP}
B.~Borchers, ``{CSDP}, a {C} library for semidefinite programming,''
  \emph{Optimization Methods and Software}, vol.~11, pp. 613 -- 623, 1999.

\end{thebibliography}
}

%\clearpage
%\appendix
%\input{doc/app_ilp}
%\input{doc/app_linear}
%\input{doc/app_graph}
%\input{doc/app_proof}

\end{document}